\documentclass{aa}
\usepackage{txfonts}
\usepackage{graphics} 
\usepackage{tabularx}
\usepackage{amssymb}
\usepackage[authoryear]{natbib}
\bibpunct[]{(}{)}{;}{a}{}{,} 

\begin{document} 

\newcommand{\mic}{\,$\mu$m }
\newcommand{\micpa}{\,$\mu$m}

\def\gtapp
{\mathrel{\hbox{\raise0.3ex\hbox{$>$}\kern-0.8em\lower0.8ex\hbox{$\sim$}}}}
\def\ltapp
{\mathrel{\hbox{\raise0.3ex\hbox{$<$}\kern-0.75em\lower0.8ex\hbox{$\sim$}}}}

\newcommand{\manque}[1]{\rule{0.5cm}{0.2cm} {\bf #1} \rule{0.5cm}{0.2cm}}
\setlength{\unitlength}{1cm}

\title{Dust enshrouded star-forming activity in Arp\,299\thanks{Based on
observations with ISO, an ESA project with
instruments funded by ESA Member States (especially the PI countries:
France, Germany, the Netherlands and the United Kingdom) with the
participation of ISAS and NASA.}}
   
\author{P. Gallais\inst{1}
        \and
        V. Charmandaris\inst{2,3}
        \and
        E. Le Floc'h\inst{1} 
        \and
        I.F. Mirabel\inst{1,4}
        \and
        M. Sauvage\inst{1} 
        \and 
        L. Vigroux\inst{1}
        \and
        O. Laurent\inst{1} 
        } 
\offprints{P. Gallais, pascal.gallais@cea.fr} 
  
\institute{ 
        Service d'Astrophysique, CEA-Saclay, F-91191, Gif sur Yvette
        Cedex, France 
        \and 
        Cornell University, Astronomy Department, Ithaca, NY 14853,
        USA
        \and 
        Chercheur Associ\'e, Observatoire de Paris, 61 Av. de
        l'Observatoire, F-75014, Paris, France
        \and 
        Instituto de Astronom\'\i a y F\'\i sica del Espacio/CONICET,
        cc 67, suc 28, 1428 Buenos Aires, Argentina
        } 
\date{Received 5 May 2003 / Accepted 18 September 2003} 
\authorrunning{Gallais et al.} 
\titlerunning{Dust enshrouded star-forming activity in Arp\,299}
 
\setlength{\unitlength}{1cm}

\abstract{
We present mid-infrared spectro-imaging ($5 - 16\,\mu$m) observations
of the infrared luminous interacting system Arp 299 (= Mrk\,171 =
IC\,694+NGC\,3690) obtained with the ISOCAM instrument aboard ISO. Our
observations show that nearly 40\,\% of the total emission at 7 and
$15\,\mu$m is diffuse, originating from the interacting disks of the
galaxies.  Moreover, they indicate the presence of large amounts of hot
dust in the main infrared sources of the system and large extinctions
toward the nuclei.  While the observed spectra have an overall similar
shape, mainly composed of Unidentified Infrared Bands (UIB) in the
short wavelength domain, a strong continuum at $\sim 13\,\mu$m and a
deep silicate absorption band at $10\,\mu$m, their differences reveal
the varying physical conditions of each component.  For each source,
the spectral energy distribution (SED) can be reproduced by a linear
combination of a UIB ``canonical'' spectral template and a hot dust
continuum due to a $230-300$\,K black body, after independently
applying an extinction correction to both of them.  We find that the
UIB extinction does not vary much throughout the system ($A_{\rm V}
\ltapp 5$\,mag) suggesting that most UIBs originate from less
enshrouded regions.  IC\,694 appears to dominate the infrared emission
of the system and our observations support the interpretation of a
deeply embedded nuclear starburst located behind an absorption of
about 40 magnitudes.  
The central region of NGC\,3690 displays a hard 
radiation field characterized by a [Ne{\sc iii}]/[Ne{\sc ii}]
ratio $\geq 1.8$. It also hosts a strong continuum from 5 to $16\,\mu$m
which can be explained as thermal emission from a
deeply embedded ($A_{\rm V}\sim60$\,mag) compact source, consistent
with the mid-infrared signature of an active galactic nucleus (AGN), and in
agreement with recent X-ray findings.
 } \maketitle
  
\section{Introduction} 

At a distance of 41\,Mpc ($v_{\rm hel} = 3080\,{\rm km\,s}^{-1}$,
$H_{0} = 75\,{\rm km\,s}^{-1}\,{\rm Mpc}^{-1}$), Arp\,299 (Mrk\,171 =
IC\,694 + NGC\,3690) is one of the nearest interacting galaxies. 
Because of its proximity, its spatial extent (8\,kpc), and its high
infrared (IR) luminosity\footnote{We use the standard definition of
$L_{\rm IR}(8-1000\,\mu{\rm m}) = 5.62\times10^5\,D^2_{\rm
Mpc}\,(13.48f_{12} + 5.16f_{25} + 2.58f_{60} + f_{100})\,L_{\sun}$
\citep[see][]{Sanders96}.} ($L_{\rm IR} = 5.16\times
10^{11}\,L_\odot$), it is one of the prime candidates for
exploring the effects of triggered star formation activity as the
tidal forces during an interaction lead to instabilities in the
galactic disks and rapidly funnel large quantities of gas into the
dynamical centers of the galaxies \citep{Mihos96,Sanders96}.  Similar to the
prototypical interacting pair of galaxies NGC\,4038/39, this system is
dynamically young, as shown by the prominent 180\,kpc
($13^\prime$) long tidal tail \citep{Hibbard99}.

\begin{figure*}[!ht]
\IfFileExists{pgallais_3926_f1.eps}{
\resizebox{\hsize}{!}{\includegraphics{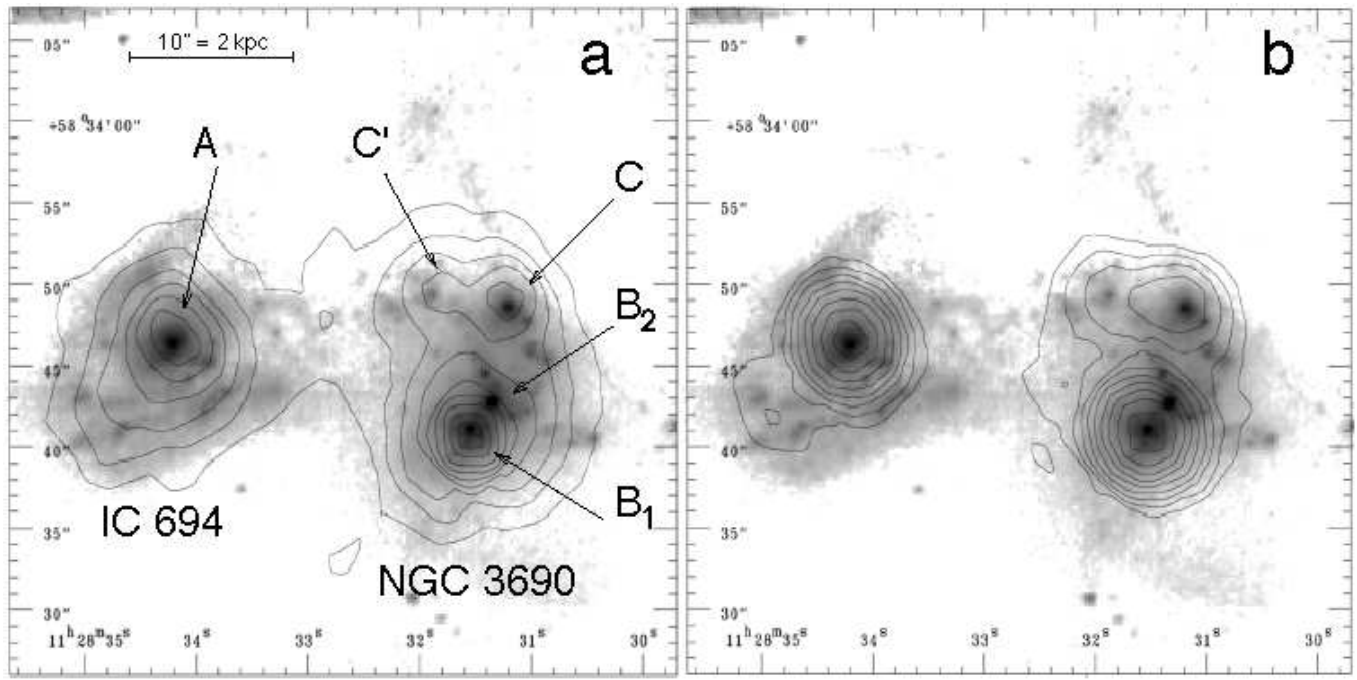}}}{\framebox(13.5,6.7){}}
\caption[bidon]{{\bf a)} The HST/NICMOS 2.2\,$\mu$m image of Arp\,299 from
\citet{Alonso00} with the overlay of the ISOCAM 7\,$\mu$m emission. 
The different components of the system are indicated.  The 9 contour
levels are set with logarithmic spacing between 1 and
33\,mJy\,arcsec$^{-2}$.  The image scale is displayed with a bar of
2\,kpc (10{\arcsec}) at the top.  {\bf b)} Same as in {\bf a)} but
using the ISOCAM 15\,$\mu$m image as an overlay, having set the
contour limits to 6 and 60\,mJy\,arcsec$^{-2}$.  Adapted from
\citet{Charmandaris02}.}
\label{fig:nicmos_CAM}
\end{figure*}

Arp\,299 has been studied extensively at all wavelengths over the past
thirty years \citep[see the original work by][]{Weedman72}.  Most
notably though, the mid-infrared and radio work of \citet{Gehrz83},
the observations in the optical by \citet{Augarde85}, in the
near-infrared by \citet{Gallais91} and \citet{Wynn91}, along with the
information on its molecular gas kinematics by \citet{Aalto97} and
\citet{Casoli99} have provided ample evidence for the extreme
characteristics of this system.  More recently, the high spatial
resolution images obtained with the Hubble Space Telescope (HST) in
the near-IR \citep{Alonso00}, as well as in the mid-IR using Keck
\citep{Soifer01} have revealed numerous point-like sources that were
undetected in the optical, and firmed up our understanding of the
complex star-forming history of the galaxy \citep[see][, and
references therein]{Alonso00}.  Following the notation suggested by
\citet{Gehrz83}, the system is usually described by the galaxy IC\,694
(source\,A) to the east, the nuclear region of NGC\,3690 (source\,B)
to the south-west and two regions of emission (sources\,C and
C$^\prime$) to the north-west.  Source\,B in NGC\,3690 has been
further resolved into several components, among which B1 and B2
clearly dominate the global emission in the near- and mid-IR (see
Fig.\,\ref{fig:nicmos_CAM} for details).  Interestingly it was also
recently shown by \citet{Charmandaris02} that despite its
inconspicuous appearance in the optical, IC\,694 contributes nearly
half of the infrared ($8-1000\,\mu$m) luminosity of the system.

One of the recurrent questions though in all studies of Arp\,299 is
whether this galaxy is a pure starburst as was proposed by
\citet{Weedman72} or whether some luminosity could be contributed by
Active Galactic Nucleus (AGN).  
The fact that it is one of the most
X-ray luminous galaxies
\citep[$L_{0.2-4.0\,{\rm keV}}\sim 10^{42}\,{\rm
erg\,s}^{-1}$, see][]{Fabbiano92,Zezas98} along with the flatness of
its radio continuum emission had given some indications that an
enshrouded AGN could be lurking in one of the nuclei.  Only recently
though the hard X-ray ($10-40$\,keV) Beppo-SAX observations of
\citet{Dellaceca02} have provided unambiguous evidence that a deeply
buried ($N_{\rm H} = 2.5 \times 10^{24}\,{\rm cm}^{-2}$) AGN is
located in one of the interacting components of the system. 
Furthermore, the same authors also demonstrated that the intrinsic AGN
luminosity in both the UV and X-rays is almost an order of magnitude
less than the infrared emission of the system calculated by
\citet{Charmandaris02}, indicating that the bulk of the infrared
emission of the galaxy is indeed due to massive star formation as
proposed by \citet{Laurent00}.  The poor spatial resolution of the
Beppo-SAX can not pinpoint the exact location of the AGN. However,
preliminary analysis of the Chandra data \citep{Zezas03},
 which provide sub-arcsecond resolution, suggests that
the AGN might be located within source B1 of NGC\,3690.

Motivated by this recent activity, we decided to analyze in detail our
$5-16\,\mu$m ISOCAM \citep{Cesarsky96} spectrophotometric data on this
system.  Our observations provide good spatial resolution with
unprecedent sensitivity and enable us to examine in detail the dust
properties of the obscured nuclear regions of Arp\,299.  The
observations and data reduction methods are described in Sect.\,2, a
brief description of the global mid-IR morphology of the system is
presented in Sect.\,3 and the analysis of the mid-IR spectra is shown
in Sect.\,4.  In Sect.\,5, we discuss the implications of our findings
on the physical characteristics of the various infrared sources and
our conclusions are summarized in Sect.\,6.

\section{Observations and Data Reduction}   

Arp\,299 was part of the ISO guaranteed time program CAMACTIV (PI.
I.F. Mirabel) which had, as a prime goal, the study of the mid-IR
properties of more than 20 nearby active/interacting galaxies
\citep{Laurent00}.  It has been observed with ISOCAM with the
Continuously Variable Filter (CVF), resulting in a full coverage of
the spectral range from 5 to 16\,$\mu$m with a spectral resolution
$R \sim 40$.  The pixel size was 1.5{\arcsec} giving a total field of
view of $48\arcsec\times 48\arcsec$.  The system was then fully mapped
with an effective spatial resolution ranging from 3 to 4.5{\arcsec}. 
The data reduction and analysis were performed using the CAM
Interactive Software (CIA)\footnote{The ISOCAM data presented in this
paper were analyzed using ``CIA'', a joint development by the ESA
Astrophysics Division and the ISOCAM Consortium led by the ISOCAM PI,
C. Cesarsky, Direction des Sciences de la Mati\`ere, C.E.A., France.}. 
Dark subtraction was done using a model of the secular evolution of
ISOCAM's dark current.  Cosmic rays were removed using a
multi-resolution median filtering method while the memory effects of
the detector were corrected using the so-called IAS transient
correction algorithm which is based on an inversion method
\citep{Abergel96}.  The flat field correction was performed using the
library of calibration data.  Finally, individual exposures were
combined using shift techniques in order to correct the effect of
jittering due to the satellite motions (amplitude $\sim 1\arcsec$). 
These methods and their consequences are discussed in detail in
\citet{Starck99}.  We estimate that the uncertainty of our mid-IR
photometric measurements is $\sim 20$\,\%.

Experience with past ISOCAM observations has indicated that the
overall pointing of ISO in the CVF mode is reliable.  However, given
the disturbed morphology of the system, we verified the relative
astrometry by comparing and identifying the main features of our
mid-IR images with those detected at other wavelengths after taking
into account the shape of their spectral energy distribution (SED) as
well as possible effects of absorption.  For this purpose, we made an
extensive use of the HST WFPC2 and NICMOS data of \citet{Alonso00}, as
well as the Keck 12\,$\mu$m images obtained by \citet{Soifer01}. 
Although these mid-IR data are not as deep as our ISOCAM observations,
their much higher spatial resolution clearly facilitated our efforts. 
Since one can not blindly rely on the astrometry information provided
by the ISO telemetry, we based our identification on the nucleus of
IC\,694 (source A), which is sufficiently isolated and its position is
well defined in both the HST near-IR images as well as in the Keck
12\,$\mu$m maps.  Finally the stability of the roll angle of the
telescope during the observations and the apriori knowledge of the
separation of the brighter mid-IR point sources enabled us to securely
identify the position of the remaining components.

\section{The Mid-IR morphology}

As mentioned earlier, the comparison between our data and the high
resolution Keck images of \citet{Soifer01} allowed us to clearly
locate the four components A, B, C and C$^\prime$ as the dominant
sources of emission in our ISOCAM images.  These identifications
further revealed that the centroid of our source B actually
corresponds the position of B1.  Even though the angular separation
between B1 and B2 is only $\sim 2\arcsec$, well below our spatial
resolution, the fact that B2 is only marginally detected in the maps
of \citet{Soifer01} leads us to believe that its contribution to the
mid-IR emission of B is minimal longward of 7\,$\mu$m.  Thus the
conclusions we will draw for source B actually reflect the
characteristics of B1.

In an attempt to increase our sensitivity to faint emission features,
we used the CVF dataset to construct broad band images at the two most
commonly used ISOCAM filters centered at 7 and 15\,$\mu$m.  The
resulting $5-8.5\,\mu$m and $12-18\,\mu$m images, which correspond to
the LW2 and LW3 filters, reach a 1$\sigma$-rms noise of 0.27 and 0.44
mJy\,arcsec$^{-2}$ respectively.  They are presented in
Fig.\,\ref{fig:nicmos_CAM}, adapted from \citet{Charmandaris02}, with
their contours overlaid on the NICMOS HST data obtained by
\citet{Alonso00}.  The LW2 image is dominated by the contribution of
the Unidentified Infrared Band (UIB) features and as a result traces
mostly the contribution of photo-dissociation and quiescent
star-forming regions.  The LW3 image, on the other hand, mainly
samples the thermal signature of hot dust from massive star-forming
regions as well as emission from high excitation forbidden lines such
as [Ne{\sc ii}]$12.8\,\mu$m and [Ne{\sc iii}]$15.6\,\mu$m.  A more
detailed description on how the underlying mid-IR SED affects the LW2
and LW3 colors in active galaxies is presented in \citet{Laurent00}. 
The mid-IR fluxes of the various sources are reported in
Table\,\ref{tab:fluxes}.

\begin{figure*}[!ht]
\IfFileExists{pgallais_3926_f2.eps}{
\resizebox{\hsize}{!}{\includegraphics{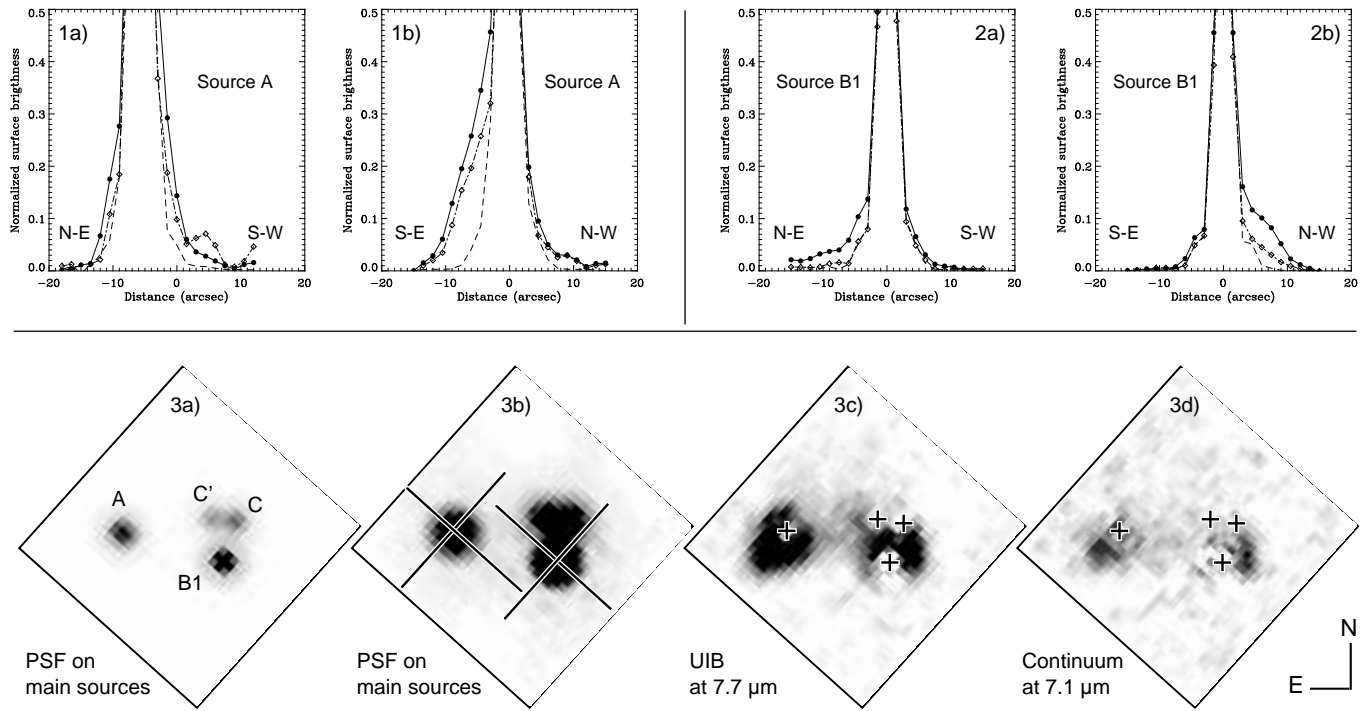}}}{\framebox(13.5,7.1){}}
\caption[bidon]{Normalized surface brigthness profiles of sources A
(panels 1a, 1b) and B1 (panels 2a, 2b) in the continuum at 7.1\,$\mu$m
(dot-dashed line with diamonds) and in the UIB band at 7.7\,$\mu$m
(filled dots on continuous line) overplotted to the normalized profile
of the PSF (dashed line) along the 2 detector axis (considered pixels
are tagged by a line in image 3b).  Image 3a and 3b represent synthetic
image built with adjusted PSFs positioned at the locations of A, B1,
C and C$^\prime$ (3a: full dynamical range; 3b: 5\% of the
maximum). Images 3c and 3d display the extended emission respectively 
in the 7.7\,$\mu$m UIB and in the 7.1\,$\mu$m continuum, i.e. the 
residual emission after subtraction of the synthetic image of PSFs to the 
image of the system in both bands (same dynamical range than the one applied 
for image 3b)}
\label{fig:psfplot}
\end{figure*}

In addition to the four main components (A, B, C and C$^\prime$)
detected in our mid-IR images, we note that there is considerable
diffuse emission, in particular at 7\,$\mu$m, in the region between
IC\,694 and NGC\,3690 as well as around the nuclei of the galaxies. 
Our equivalent broad band images reveal that this diffuse component,
which was undetected in the shallower ground based images, accounts
for 44\,\% and 36\,\% of the total 7 and 15\,$\mu$m fluxes originating
from Arp\,299.  
This diffuse emission is shown in Fig.\,\ref{fig:psfplot} which 
displays normalized profiles at 7.1\,$\mu$m (continuum) and 
7.7\,$\mu$m (UIB band) of sources A and B1 compared to the PSF profiles. 
Together with the profiles, images of the diffuse emission at these 
wavelengths are presented, obtained by subtraction of a synthetic image 
made of
PSFs positioned at the locations of sources A, B1, C and C$^\prime$ to 
the images of the system in the 7.7\,$\mu$m UIB and in its continuum. If 
we derive an equivalent IRAS 12\,$\mu$m flux
assuming a power-law SED in the $5-18\,\mu$m range, we find that
60\,\% of the overall IRAS 12\,$\mu$m flux is accounted for by the
four sources, with A and B1 being the dominant ones (22 and 28\,\% of
the total flux respectively).

\begin{table}[htb]
\caption[bidon]{Broad-band mid-IR photometry of Arp\,299}
\label{tab:fluxes}
\begin{tabular}{lccc}
\hline 
\hline
Source & LW2$^{\dag}$ (mJy)   &  LW3$^{\dag}$   (mJy)  &  LW3/LW2
\\
\hline
A            &  325 &  1860 & 5.72 \\
B            &  505 &  1951 & 3.86 \\
C            &  126 &   461 & 3.66 \\
C$^\prime$   &   76 &   232 & 3.05 \\
Total        & 1846 &  7037 & 3.81 \\
\hline
\end{tabular}
\noindent\\
$^{\dag}$ using a beam of 4.5{\arcsec} in diameter 
(aperture corrections were applied to account for the
overall extension of the PSF).
\end{table}

All main components in the system are very bright in both filters and
have extremely red colors.  The 15 to 7\,$\mu$m flux ratios LW3/LW2
range from 3 to 5.7 and they are quite high compared with what is
typically found in normal galaxies where they vary from 0.7 to 1.2
\citep{Roussel01}.  Even in the case of the Antennae galaxies, Knot A,
the brightest 15\,$\mu$m region, exhibits a LW3/LW2 ratio of 2.6
\citep{Mirabel98}, while the same indicator in the massive
star-forming complex within the outer ring of the Cartwheel galaxy is
5.2 \citep{Charmandaris99}.  In the absence of an AGN-type activity,
such high ratios in Arp\,299 indicate a star-forming mechanism of a
remarkably high efficiency \citep{Gallais99,Laurent00}, a result which
has also been suggested by \citet{Lai99} and \citet{Rouan99} based on
their near-IR adaptive optics observations.  Using a gray body model
to fit the mid- and far-IR SED of the system, \citet{Charmandaris02}
have estimated that the infrared luminosities of the three main
components (A, B1+B2 and C+C$^\prime$) range from 0.44 to $1.8\times
10^{11}\,L_\odot$.  This implies that each of these sources is several
times more luminous than most starburst galaxies observed in the local
Universe such as the prototypical M\,82.

Examining the individual sources in more detail, it is striking that,
within the limits of our resolution, component\,A is dominated in the
mid-IR by a strong nuclear source with a faint extension towards the
southeast.  This nuclear source is completely absent from the optical
HST images and only becomes apparent in the original near-IR maps of
\citet{Wynn91}.  In the near-IR, this is interpreted as the emission
from the old stellar population in the bulge of IC\,694.  However, the
mid-IR spectrum (see Fig.\,\ref{fig:spectres}) differs clearly from
the one of a population of old stars, such as in elliptical galaxy, in
which there is no emission or absorption band \citep{Athey02}.  It is
rather similar to the one we encounter in star forming regions
\citep{Laurent00}. In the mid-IR, the dominant process is star
formation 
which is not evidenced at shorter wavelengths
(visible and near-IR) and it appears mainly concentrated in the
nuclear region.
At longer wavelengths, the emission of A
dominates the global infrared emission of the system, making A the most
luminous component of the system at $38\,\mu$m \citep{Charmandaris02}. 
This implies that even though star formation does happen over an
extended area of IC\,694 -- as evidenced by the dust lanes and star
forming tidal streamers visible in the HST optical imagery of
\citet{Malkan98} -- and does contribute in the mid-IR by creating an
extended emission seen in Fig.\,\ref{fig:nicmos_CAM}, the bulk of the
infrared flux originates from its nucleus.

\begin{figure}[!ht]
\IfFileExists{pgallais_3926_f3.eps}{
\resizebox{\hsize}{!}{\includegraphics{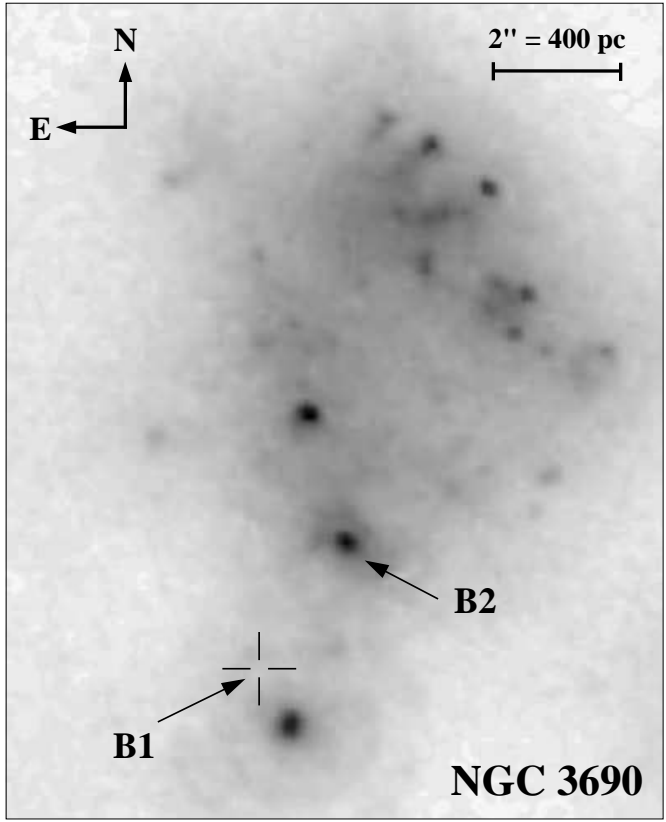}}}{\framebox(13.5,16.8){}}
\caption[bidon]{UV image (220\,nm) of NGC\,3690 taken with the FOC on-board
the HST (see also \citet{Meurer95}) with 
positions of B1 
and B2
marked.  One can see that no emission at this wavelength is associated
with B1 confirming its highly obscured nature.}
\label{fig:foc}
\end{figure}

The mid-IR structure of NGC\,3690 is more complex.  As discussed in
detail by \citet{Alonso00}, a large number of point-like sources can
be seen in the near-IR, with B1 and B2 largely dominating the overall
emission of the galaxy at these wavelengths.  The spectral properties
of these two main components though present a remarkable contrast.  As
the wavelength increases, B1 becomes progressively brighter while the
contribution of B2 -- clearly the most luminous source of NGC\,3690 in
the visible and in the J and H bands \citep{Alonso00} -- gently
declines longward of 1.6\,$\mu$m.  Furthermore, the latter is barely
detectable in the high resolution mid-IR images of \citet{Soifer01},
while B1 clearly unveils a very luminous core at 12\,$\mu$m.  The red
colors of B1 and its differential extinction of $A_{\rm V}=15$\,mag
relative to B2 \citep{Alonso00} are indicative of its much more
obscured nature and reveal the presence of large concentrations of
dust and gas in this area.  As a side note, it is worth mentioning
that B1 is completely invisible in the HST/FOC 220\,nm UV map of the
galaxy (see Fig.\,\ref{fig:foc}), and it is very faint at 300\,nm
\citep[see][Fig.\,3.34a,b]{Windhorst02}, an additional evidence of its
high dust content.

Sources C and C$^\prime$ are clearly resolved at 7\,$\mu$m but the
emission from C$^\prime$ diminishes relatively to C at longer
wavelengths (see Fig.\,\ref{fig:nicmos_CAM}).  It has been proposed
that source C could be the nucleus of a third galaxy in the
interacting system, but this hypothesis is questioned by the location
of this source on the [J-H]/[H-K] color-color diagram
\citep{Gallais91}, far from the locus of the other nuclei.  Recent CO
observations \citep{Aalto97,Casoli99} also suggest that C and
C$^\prime$ are dynamically linked and can be considered as
star-forming complexes in the overlap region of NGC\,3690 and IC\,694.

\section{Mid-IR spectral properties and absorption}
\label{section:spectro}

\begin{figure*}[!ht]
\IfFileExists{pgallais_3926_f4.eps}{
\resizebox{\hsize}{!}{\includegraphics{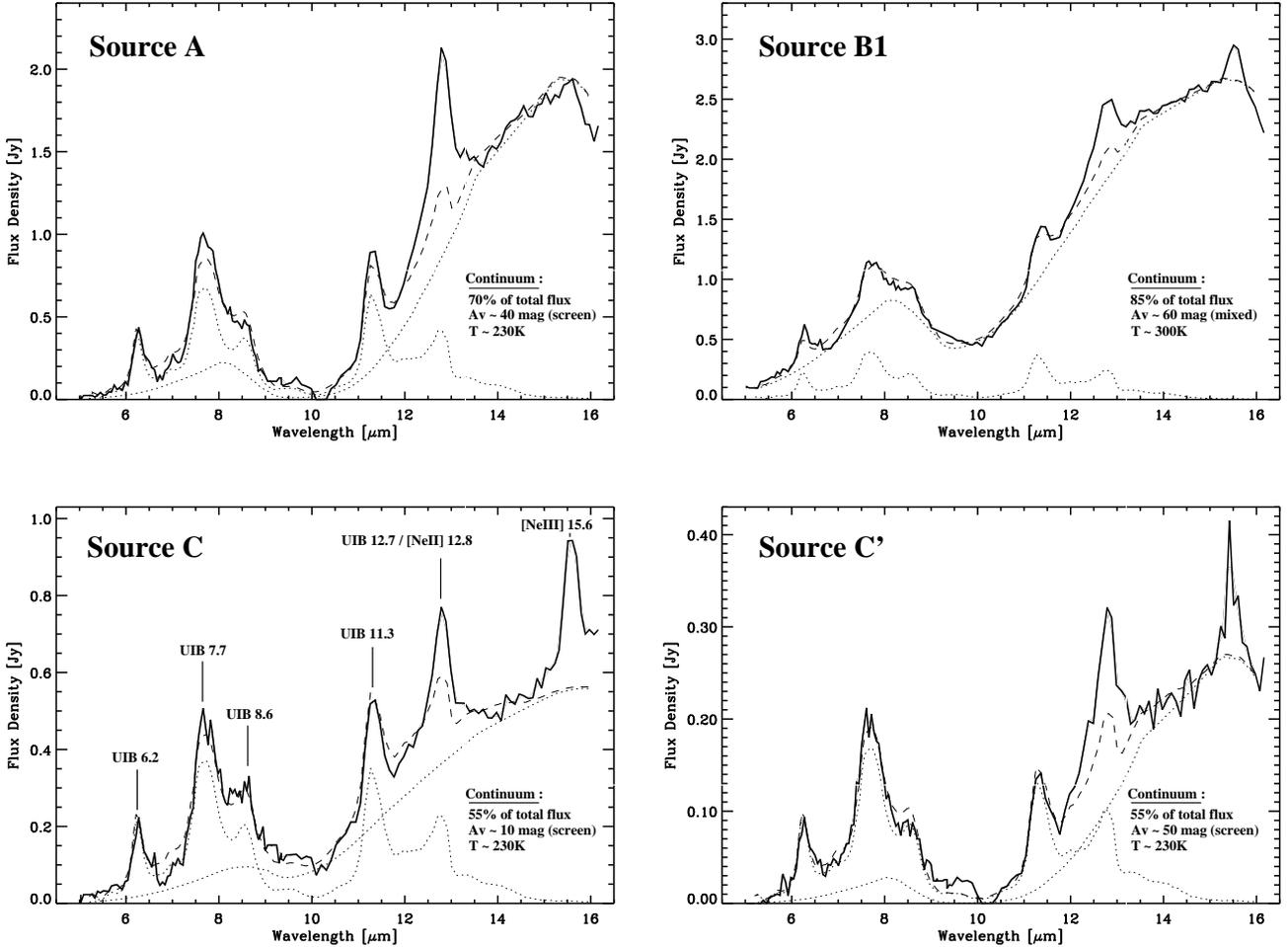}}}{\framebox(13.5,9.8){}}
\caption[bidon]{Rest-frame mid-infrared spectra of the four main sources (A,
B1, C and C$^\prime$) in Arp\,299 ({\it solid lines}\,).  They were
obtained with a $4.5^{\prime\prime}\times 4.5^{\prime\prime}$ aperture
centered on each component, and they were corrected via a scaling
factor to account for the extended flux of the PSF lying outside of
the aperture.  For each component, we have also plotted our best fit
({\it dashed line}\,) obtained with the combination of a UIB template
and a black body continuum ({\it dotted lines}\,) extinguished with
the law of \citet{Dudley97}.  On each panel, we indicate the physical
properties of the continuum: its contribution to the mid-IR flux, the
absorption magnitudes followed by the dust obscuration geometry (screen
against mixed model), and the temperature of the fitting black body. 
We also found that the extinction level of the UIB component is
negligible ($A_{\rm V} \ltapp 5$\,mag) in the four sources.  Note that
the [Ne{\sc ii}]12.8\,$\mu$m and [Ne{\sc iii}]15.6\,$\mu$m emission
lines were not taken into account in our models.}
\label{fig:spectres}
\end{figure*}

In Fig.\,\ref{fig:spectres}, we present the $5-16\,\mu$m spectra of
the four main sources A, B1, C and C$^\prime$ of Arp\,299.  They have
been obtained with a $4.5^{\prime\prime}\times 4.5^{\prime\prime}$
aperture centered on the brightest spot of each component.  A
wavelength-dependent scaling factor has also been applied to correct
for the extended flux of the PSF lying outside of the aperture used. 
We observe that the overall shape of each spectrum is similar, with
the UIB features dominating the emission in the short wavelength range
and with a warm continuum becoming visible longward of 10\,$\mu$m. 
Our measurements are also consistent with the $8-13\,\mu$m
spectroscopic observations already obtained by \citet{Dudley99}.

A more thorough analysis though, reveals distinct underlying physical
properties of each region.  While the mid-IR SED of source C is
typical of what is generally observed in starburst galaxies
\citep[e.g.,][]{Laurent00}, the spectra of sources A and C$^\prime$
display very strong absorption in the 9.7\mic silicate band and a
significant contribution of the thermal continuum at $\lambda >
12\,\mu$m.  Sources C and C$^\prime$ also display a strong [Ne{\sc
iii}] emission line, which as we discuss in section 5, is not as
apparent in the other components.  Finally, we note that the spectrum
of source B1 is rather intriguing since it is dominated by an
important hot dust continuum throughout the whole CVF spectral range,
with only a minor contribution of the UIB features.

To provide a more quantitative insight into the properties of these
sources, their $5-16\,\mu$m SEDs were carefully analyzed and compared
with a library of simulated-CVF spectra.  The latter were created with
a two-component model described hereafter.  The role of the first
component is to take into account the presence of the UIB features,
usually exhibiting a canonical SED as it has been observed in various
star-forming galaxies \citep{Dale01,Roussel01}, so we used the average
$5-16\,\mu$m SED of the M\,51 galactic disk as a template spectrum. 
The second component, which reproduces the warm continuum observed in
the long wavelength part of the CVF, was modeled using a black body
with a temperature typically ranging from 150 to 300\,K. We also
considered the use of a power law with a spectral index varying from
$\alpha=1$ to $\alpha=2$, as well as black bodies modified by an
emissivity law \citep[see also][]{Sturm00,Forster03}.  However, given
the similar shape of the power law and black body functions at these
wavelengths and temperature regimes as well as the limited wavelength
coverage of our spectra, we found that we were unable to discriminate
between one of the two particular spectral shapes.  As a result, we
will restrict our analysis using simple black body fits.  Selective
reddening was then applied {\it independently} towards these two
components \citep[similarly to the approach followed by][]{Tran01},
and a variety of extinction laws \citep{Mathis90,Dudley97,Lutz99,Li01}
were explored assuming either a uniformed dust screen geometry
($F_{\lambda, out} = F_{\lambda, in} \times \exp^{-\tau_{\lambda}}$ ,
$\tau_{\lambda} = A_{\lambda}/1.086 $) or a homogeneous mixture of
dust grains and sources ($F_{\lambda, out} = F_{\lambda, in} \times (1
- \exp^{-\tau_{\lambda}}) / \tau_{\lambda}$ , $ \tau_{\lambda} =
A_{\lambda}/1.086 $).  Varying the fraction of the UIB and continuum
emission to the total mid-IR flux, a collection of CVF spectra was
subsequently created and $\chi^2$--tested against our data to select
the best simulated SED available in our library.  Note that the strong
ionic emission lines usually observed in starburst environments
\citep{Sturm00} such as [Ne{\sc ii}] at 12.8\mic and [Ne{\sc iii}] at
15.6\mic were not taken into account in our simulated CVF spectra as
they are not included in our model.  Moreover, at the spectral
resolution of our CVF, [Ne{\sc ii}] is blended with the 12.7\mic UIB
feature.  Therefore the $12-13.2\,\mu$m and $15-16\,\mu$m wavelength
ranges were omitted in our $\chi^2$ calculations.

In Fig.\,\ref{fig:spectres}, we present the best fits we obtained for
the four ISOCAM spectra, along with the decomposition into their two
components.  While the specific results which were derived for each
source will be more thoroughly discussed in the next section, we
highlight hereafter some general points of our findings, as well as a
number of limitations related to our fitting procedures. 

One of our most striking results is that our best fits indicate an
elevated extinction for the underlying continuum emission of sources
A, B1 and C$^\prime$ and a very low extinction ($A_{\rm V} \ltapp
5$\,mag) of the template which contributes to the UIB feature
emission. This may suggest not only a physically different origin but
also a different spatial distribution for the two components observed
in our spectra. Such a picture could actually be understood if we
consider that the continuum emission originates from deeply
dust-enshrouded regions while the UIBs are more diffuse and have a
widespread and surface distribution. Since the UIBs mostly trace areas
of photo-dissociation surrounding H{\sc ii} regions, this would not be
so surprising given the filamentary optical/near-IR emission
\citep{Malkan98,Alonso00} and the extended gas streamers
\citep{Sargent91,Casoli99} which have been observed in the in-between
regions of Arp\,299.

A potential limitation in our approach was revealed by the degeneracy
originating from the unknown geometry of the dust in the obscured
regions.  For all sources but B1, we found that both the screen and
the mixed models can actually lead to similarly good overall fits to
the observed spectrum in terms of the $\chi^2$ minimization, even
though the underlying assumptions for the geometry of the sources are
different.  For example, the continuum emission of source\,A can
easily be reproduced either with a 230\,K black body extinguished with
$A_{\rm V} = 40$\,mag in a screen geometry (see
Fig.\,\ref{fig:spectres}), or with a much cooler component at $T =
160\,$K with $A_{\rm V} = 140$\,mag assuming the mixed model. 
However, we wish to stress that the physical meaning of such
single-temperature fits is rather uncertain since {\it i}) the
observed spectrum is the result of no single grain population but a
linear superposition of grains with a specific size distribution
\citep[see][]{Dale01}, {\it ii}) the dust grains responsible for this
warm continuum are not in thermal equilibrium, and {\it iii}) the
presence of temperature gradients in the close vicinity of the mid-IR
sources is quite likely.  Therefore, a more meaningful approach would
rather be to compare the four fitted spectra to each other after
having primarily defined a given geometry for the dust obscuration
(see Sect.\,\ref{sec:discussion} for such comparisons).  Furthermore,
we note that both the fraction of UIBs to the total mid-IR flux and
their negligible extinction measured using our fitting procedure are
roughly independent of the screen or mixed model assumption.

\begin{table}[bth]
\caption[bidon]{
Intensities and ratio of the [Ne{\sc ii}]12.8\,$\mu$m and [Ne{\sc
iii}]15.6\,$\mu$m for Arp299.  The values of [Ne{\sc ii}] were
measured by \citet{Dudley99} and should be considered as upper limits
as they may be contaminated by the 12.7\,$\mu$m PAH feature (see
text).
}
\label{tab:neon}
\begin{tabular}{lccc}
\hline\hline
&  $I$([Ne{\sc ii}])    &    $I$([Ne{\sc iii}])  & \\
\raisebox{1.5ex}[0cm][0cm]{Source}   &   {\tiny ($10^{-15}\,{\rm 
W\,m}^{-2}$)} & {\tiny ($10^{-15}\,{\rm 
   W\,m}^{-2}$)} & 
\raisebox{1.5ex}[0cm][0cm]{\large $\frac{I([{\rm 
NeIII}])}{I([{\rm NeII}])}$}   
\\
\hline
A  &   $<2$ & -- & --\\
B1 &   $0.7$ & $>1.3$ & $\geq 1.8$\\
C  &   $2.0$ & 1.3 & $\geq 0.65$ \\
C$^\prime$ &   $<0.9$ & 0.5 & $>0.55$\\
\hline
\end{tabular}
\end{table}

To investigate the hardness of the radiation field in the different
regions of the system we can use the ratio of the ionic Neon lines
[Ne{\sc ii}]12.8\,$\mu$m and [Ne{\sc iii}]15.6\,$\mu$m.  Even though
we can easily measure the [Ne{\sc iii}] line, the spectral resolution
of the CVF ($R \sim 40$) is just too low to distinguish the [Ne{\sc
ii}]12.8\,$\mu$m line from the 12.7\,$\mu$m PAH band.  As a result we
decided to use the [Ne{\sc ii}] values of the system measured by
\citet{Dudley99} who calculated the intensity of the [Ne{\sc ii}]
lines in all sources of Arp\,299.  His $R \sim 60$ spectrograph can
almost resolve the [Ne{\sc ii}] line, and the 5.5\,{\arcsec} circular
aperture he used is only $\sim 25$\,\% smaller than the aperture used
to extract our ISOCAM/CVF spectra presented in
Fig.\,\ref{fig:spectres}.  Our results are presented in Table
\ref{tab:neon}.  In principle, the values of [Ne{\sc ii}] may still
contain some contribution from the 12.7\,$\mu$m PAH feature, thus
making the [Ne{\sc iii}]/[Ne{\sc ii}] ratio a lower limit. However, this
contribution must be small since the sum of line intensities
measured by \citet{Dudley99} in the regions B1, C, and C$^\prime$ is
consistent with the high spectral resolution ISO/SWS measurements of
\citet{Thornley00}.
We note an inconsistency between the 
sum of our measurements on B1, C and C$^\prime$ and the integrated 
measurement presented in \citet{Thornley00}. Whether their value 
underestimated or our overestimated is not easy to determine, and 
spectroscopic observations of this target with SIRTF will resolve 
this issue.

\begin{figure*}[!ht]
\IfFileExists{pgallais_3926_f5.eps}{
\resizebox{\hsize}{!}{\includegraphics{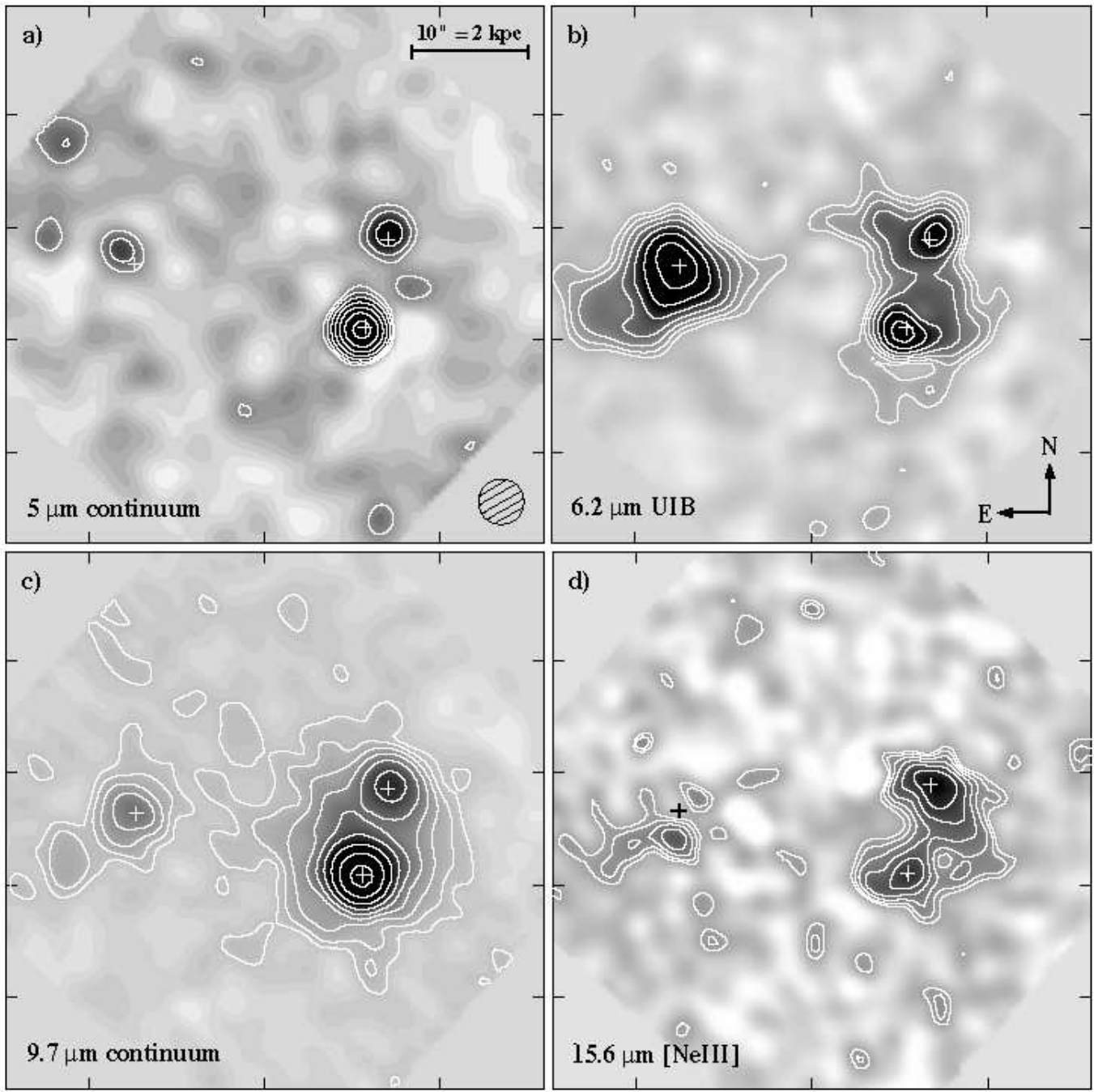}}}{\framebox(13.5,13.5){}}
\caption[bidon]{Inverted grey-scale images of Arp\,299 in selected spectral
regions, depicting {\bf a)} the hot continuum emission observed at
5\,$\mu$m, {\bf b)} the photo-dissociation regions obtained by
integrating the mid-IR emission within the 6.2\mic UIB, {\bf c)} the
continuum emission integrated over a $\Delta\lambda \sim 1\,\mu$m
bandwidth in the range of the silicate absorption at 9.7\micpa, and
{\bf d)} the distribution of the ionic [Ne{\sc iii}] emission line at
15.6\micpa, after subtraction of the underlying continuum. 
This last 
image must be considered carefully because the continuum per pixel on 
source B may have been badly estimated due to the low signal to 
noise ratio 
of the emission in the line and the probable influence of the wide $18\,\mu$m 
silicate absorption band.
For each image, the overlaid contours
have been set with logarithmic spacing.  By increasing order of
wavelengths, the minimum levels are respectively 0.44, 0.18, 0.22 and
0.49\,mJy\,arcsec$^{-2}$ while the maxima are 10.7, 2.0, 15.5 and
1.5\,mJy\,arcsec$^{-2}$.  The positions of the brightest sources (A,
B1 and C) over the $5-16\,\mu$m wavelength range are marked with
crosses.  The four images have the same orientation. The average FWHM
of the ISOCAM PSF over the full CVF spectral range ($\sim
4^{\prime\prime}$) is indicated in panel {\bf a)} as well as the scale
in arcsec and kpc.}
\label{fig:bands}
\end{figure*}

One of the great advantages of the ISOCAM/CVF observations is that we
can extract narrower-band images ($\Delta\lambda \sim 1\,\mu$m) of our
target, effectively constructing two dimensional maps of specific
spectral features.  Several maps of selected spectral regions were
created for Arp\,299 and are displayed in Fig.\,\ref{fig:bands}.  The
first is the hot dust continuum at 5\,$\mu$m, which could be due to
either thermal emission from an AGN or the Rayleigh-Jeans tail of the
photospheric emission from an old stellar population, or 
fluctuating small dust grains without aromatic features 
\citep{Helou00}. 
The integrated
emission of the 6.2\,$\mu$m UIB is also presented, since it isolates
the emission from the photo-dissociation regions of the merger and
despite the fact that it is weaker than the 7.7\,$\mu$m UIB it is much
less affected by the depth of the 9.7\,$\mu$m silicate band \citep[see
discussion in ][]{Laurent00}.  We also mapped the total flux within
the band of the silicate absorption at 9.7\,$\mu$m, and that of the
ionic [Ne{\sc iii}]15.6\,$\mu$m emission line which is indicative of
the hardness of the radiation field \citep[e.g.,][]{Thornley00}.  For
the latter and the 6.2\,$\mu$m UIB, we identified the corresponding
spectral feature within each pixel of the CVF images, and carefully
removed the underlying continuum so as to only keep the contribution
of the given emission line (see \citealt{Laurent00} and
\citealt{Lefloch01} for an illustration of the method).  
The dominant
sources of mid-IR emission (A, B and C) are clearly apparent in all
maps, even though their relative strength varies at each wavelength.
However, the 
image of the ionic [Ne{\sc iii}]15.6\,$\mu$m line remains indicative
of 
the spatial distribution of this feature. The spectra of individual 
pixels are noisier than any integrated spectra and the determination 
of the continuum pixel-per-pixel can be altered by this noise. Moreover, 
especially in the case of source B1, the depression in the SED over 
$16\,\mu$m can lead to a bad estimation of the continuum and then 
to a bad determination of the flux in the line. So, integrated spectra 
on each main source are more reliable in term of relative flux.
Even if C appears as the dominant source in the [Ne{\sc 
iii}]15.6\,$\mu$m image of Fig.\,\ref{fig:bands}.d, B1 remains the 
brighest source in this line.

\section{Discussion}
\label{sec:discussion}

\subsection{The nuclear region of IC\,694}

In our mid-IR images, IC\,694 (source A) appears as a bright source
corresponding to the radio peak observed at 6 and 20\,cm by
\citet{Gehrz83}.  As pointed out \cite[see][ and references
therein]{Alonso00,Charmandaris02}, unlike the infrared view, images in
the UV and visible do not reveal any clear optical counterpart at this
position, while large quantities of high density molecular gas are
concentrated within the central 500\,pc \citep{Sargent91,Casoli99}. 
This gas fuels a massive starburst as evidenced by a supernova rate of
0.65 yr$^{-1}$, nearly five times the combined rate observed in B and
C \citep{Alonso00}.  This large gas reservoir can easily explain the
high extinction toward the nucleus.  In the optical, a dust lane runs
in the southeast/northwest direction and ``patchy'' emission from
surface star forming regions is seen in an extended area around the
nucleus \citep{Malkan98}.  Not surprisingly, those same regions
contribute to the 6.2\,$\mu$m UIB emission and, as we observe in
Fig.\,\ref{fig:bands}b, the angular size of source A is considerably
larger at this wavelength (see also Fig.\,\ref{fig:psfplot} 
which shows this diffuse emission).

Our mid-IR spectrum of this source (see Fig.\,\ref{fig:spectres})
reveals a strong continuum with a slope past 10\,$\mu$m steeper than
in any other component of the system.  Moreover, the $9.7\,\mu$m
silicate band appears saturated.  The continuum emission from A can be
fitted with a 230\,K black body absorbed by $A_{\rm V} \sim 40$\,mag
assuming a uniform screen of dust.  We also note that the longer
wavelength part of the spectrum seems to decrease again past
15.5\,$\mu$m.  The same behavior can be seen in source B. Since we do
know from multiple observations between 20 and 38\,$\mu$m that the
spectrum of A rises quickly making it the dominant source of the IR
luminosity of the system \citep{Charmandaris02}, it is reasonable to
attribute this brief decrease in flux to the presence of the
18\,$\mu$m silicate feature.  Indeed, as shown by \citet{Dudley97},
the two silicate bands are proportional in strength/depth so, given
the observed shape of the 9.7\,$\mu$m feature, 
the 18\,$\mu$m one
would have to be strong too.
Unfortunately given the limited
wavelength coverage of the CVF and the fact that memory effects of the
detector are more pronounced at the longer wavelengths which are
observed at the beginning of the scan, we can not elaborate more on
its actual strength.  We will have to wait for the Infrared
Spectrograph (IRS) aboard the Space Infrared Telescope Facility
(SIRTF) which will provide complete wavelength coverage up to
40\,$\mu$m to address this issue more accurately.

A unique aspect of IC\,694 that separates it from all other areas in
the galaxy is the weak [Ne{\sc iii}]15.6\,$\mu$m emission. As we can
see from Fig. \ref{fig:spectres}a, even though a suspicious feature at
$15.6\,\mu$m could be the [Ne{\sc iii}] line, its equivalent width is
very small and makes it practically impossible to measure with our
resolution. This can not be easily explained by advocating that the
blue wing of the 18\,$\mu$m silicate feature extends to the location
of [Ne{\sc iii}] reducing the line flux. If simple extinction was the
culprit then how could one detect the line in both regions B1 and
C$^\prime$ where even higher values of $A_{\rm V}$ are found?
Furthermore, why does it seem that the spatial distribution of the
faint [Ne{\sc iii}] avoids the nucleus but is associated with the
dusty filaments to the southeast of source A?

The fact that the mid-IR neon lines of IC\,694 have also been observed
with ISO/SWS with superior spectral resolution by \citet{Thornley00}
can help us explore these questions.  Using a considerably larger
aperture ($14\,{\arcsec}\times 27\,{\arcsec}$) these authors easily
detected both [Ne{\sc ii}]12.8\mic and [Ne{\sc iii}]15.6\,$\mu$m
toward IC\,694.  They measured a [Ne{\sc iii}]/[Ne{\sc ii}] ratio of
0.29, a factor $\sim 2.5$ times lower than what was found for the
whole of NGC\,3690 (combined sources B and C).  Modeling by
\citet{Thornley00} has shown that low values of the [Ne{\sc
iii}]/[Ne{\sc ii}] ratio can be explained in terms of a larger age of
the starburst, lower limit in the upper mass function cut off, higher
metallicity, or a reduced ionization parameter $U$.  Our observations
are actually consistent with all these possibilities if we consider
that IC\,694 contains the larger fraction of molecular gas and cold
dust and dominates the IR luminosity of Arp\,299.  This implies that
the starburst in source A has been taking place for a longer period of
time than in sources B and C since the former seems to be closer to
dynamical equilibrium than the two sources in NGC\,3690 which are even
more disturbed and obviously still strongly interacting
\citep{Casoli99}.  One could further speculate that the geometry in A
is different from the single cluster case of \citet{Thornley00}.  The
spatial extent associated with the age of the starburst indicates that
the distribution of the enshrouded stars is more random forming
clusters resembling the mix gas/star cluster case.  
Furthermore, more
generations of stars must have been created in IC\,694, and passing
through their AGB phase, they produced the dust we observe and also lead 
to a higher metallicity in the region.
As we will discuss in the following
section this is not the case for C$^\prime$ which must harbor the
youngest OB stars in the system \citep{Soifer01}.

If the above scenario is correct then the reason why we detect [Ne{\sc
iii}] in the southeast outskirts of source A, and not in the center,
is simply due to the fact that these are the regions where young stars
form along the tidal filaments \citep{Malkan98}.  In these filaments,
the geometry is much more favorable so that ``bubbles'' in the
interstellar medium form easier exposing the high ionization radiation
of some underlying recently formed massive stars.

\subsection{On the mid-IR properties of sources C and C$^\prime$}

The shape of the mid-IR SED of these regions (see
Fig.\,\ref{fig:spectres}) is consistent with those of star-forming
regions in other extragalactic sources, where a hot thermal continuum
strongly affects the PAH emission.  Furthermore, the values of the
mid-IR diagnostic flux ratios LW3/LW2 (see Table \ref{tab:fluxes}) are
similar to those found in regions dominated by young massive stars
\citep[see][]{Laurent00}.  As a result, our data provide further
support to the conclusions of other authors \cite[][\,and references
therein]{Soifer01} that C and C$^\prime$ host extraordinary
extra-nuclear starbursts.

Although it is probable that some emission from C could contribute to
C$^\prime$ because of their small angular separation, the fact that
their measured absorptions are different indicates this is not a very
likely event. Our modeling of the extinction suggests that the
environments for both sources C and C$^\prime$ are similar, with a
higher extinction toward C$^\prime$.  For both, as for source A, the
observed spectrum can be reproduced by the simple superposition of a
UIB contribution and a black body heated to $\sim 230$\,K in order to
account for the hot dust continuum which dominates above $10\,\mu$m.
The estimated extinction for sources C and C$^\prime$ is $A_{\rm V}$
of 10 and 50\,mag respectively assuming a foreground screen geometry.

The hardness of the UV radiation field in these sources can be
characterized by the [Ne{\sc iii}]/[Ne{\sc ii}] ratio.  Source C
presents a ratio of $\geq 0.65$ and C$^\prime$ a ratio $\geq
0.55$\footnote{We must be careful considering [Ne{\sc iii}] in source
C$^\prime$.  We notice that the emission in this line is only
marginally mapped at the C$^\prime$ position (see
Fig.\,\ref{fig:bands}).  Because C and C$^\prime$ are separated only
by $5^{\prime\prime}$, the feature we observe in the C$^\prime$
spectrum may be a relic of the extended wings of the PSF (FWHM $\sim
4.5^{\prime\prime}$) centered on source C.}, both greater than the one
measured in component A \citep[0.29 from ][]{Thornley00}.  This would
indicate that the UV radiation field, and consequently the fraction of
young massive stars is higher in C and C$^\prime$ than in A, leading
to younger starbursts in these regions, in agreement with the
conclusions of \citet{Soifer01} as well as of \citet{Alonso00} who
derived $\sim 5\,$Myr, 4\,Myr and $\sim 11\,$Myr for C, C$^\prime$ and
A respectively (based on an evolutionary starburst model using
parameters derived from their observations).

\subsection{The nature of source B: Is there mid-IR evidence for an AGN?}

As we mentioned earlier, B1 appears to be the brightest source of the
system in the 5 to $16\,\mu$m spectral range and remains unresolved by
our observations.  While our spatial resolution is not sufficient to
unambiguously resolve B1 and B2, the fact that the centroid of the
source located at B position coincides very well with the 6\,cm radio
source observed by \citet{Gehrz83} indicates reasonably that we
actually map mainly the emission coming from B1.  This is in agreement
with higher spatial resolution ground-based observations from
\citet{Soifer01} who resolved B1 and B2 but showed that, despite the
fact that B2 dominates the flux below $2\,\mu$m \citep[see
also][]{Wynn91,Gallais91}, its emission decreases with increasing
wavelength and it is actually $\sim 35$ times fainter than B1 at
$12.5\,\mu$m.  Consequently, all our conclusions on the mid-IR
characteristics of source B=B1+B2 would actually reflect the
properties of B1.

Until recently, most researchers had focussed on the starburst
properties of this region, given the absence of any evidence for an
active nucleus.  Based on the H$\alpha$ \citep[see Fig.\,3
of][]{Alonso00} and CO observations \citep{Casoli99, Aalto97}, it is
clear that B1 contains a large amount of ionized gas, as well as cold
molecular gas which could sustain massive star formation activity for
more than $10^8$ years.  Our observations though indicate that the
physical characteristics of B1, as displayed by its spectrum, make it
stand out in the mid-IR in two ways.

First, B1 displays the strongest continuum emission over the whole
$5-16\,\mu$m range with only a weak contribution due to PAH emission
(see Fig.\,\ref{fig:spectres}).  Using our model, we find that this
continuum can be best fitted with a 300\,K black body extinguished by
$A_{\rm V} \sim 60\,$mag in a mixed-model geometry and accounts for
85\,\% of the total mid-IR emission of B1.  Despite the high
extinction derived from our model, B1 remains the brightest source of
the system in the $9.7\,\mu$m silicate absorption band and its ``hot
continuum'' at $\sim 5\,\mu$m is 118\,mJy, almost 5 times higher than
source A which dominates the global IR luminosity of the whole system.

Second, the [Ne{\sc iii}]/[Ne{\sc ii}] ratio of this source 
is at
least 1.8 (see Table \ref{tab:neon}),
a value considerably greater
than the ratios measured in the \citet{Thornley00} starburst sample
which typically ranges from 0.05 to 1, with the only exception of the
low metallicity systems NGC\,5253 and II\,Zw\,40 (with values of 3 and
12 respectively).  Clearly, the radiative environment in B1 deviates
from the one seen in classical starburst galaxies, and it appears to
resemble more what is observed in young galactic H{\sc ii} regions
such as W\,51 \citep{Thornley00}. \citet{Sturm02} reports measurements 
in the Neon lines for Seyfert galaxies which clearly indicate a 
higher ratio in pure Seyfert galaxies (usually $> 1$ 
with a maximum of 2.8 in their sample) than in mixed 
Seyfert/starburst systems ($<0.6$ typically). 
On the other hand, 
the ratio [Ne{\sc iii}]/[O{\sc iv}] is usually low ($< 1.5$) in Seyfert 
galaxies as computed from values reported by \citet{Sturm02} while it is higher 
in starburst galaxies, ranging from 5 to $> 30$ \citep[see][]{Verma03}. 
Assuming that C and C$^\prime$ contributes to the flux measured in [O{\sc 
iv}] by \citet{Verma03}, B1 then presents a ratio 
[Ne{\sc iii}]/[O{\sc iv}] 
$> 13$. Even if those ratios are not criteria used to discriminate 
Seyfert and starburst galaxies, the association of a high 
[Ne{\sc iii}]/[Ne{\sc ii}] with a high [Ne{\sc iii}]/[O{\sc iv}] in 
source B1 makes it of very intriguing origin.

The fact that no UV emission is detected from B1 is further evidence
of the high obscuration in the region.  In this context, the PAH
features could be attributed to individual heating sources distributed
around B1, in a diffuse component on the line of sight, in such a way
that the filling factor of their corresponding Stromgren spheres is
high and hence there is a smaller volume available for the
photo-dissociation regions to contribute to the PAH flux \citep[see][,
Fig.\,7b]{Alonso00}.

However, hard UV photons can also be produced by an AGN. Based on a
diagnostic diagram which compares the ratio of the continuum at
$14-15\,\mu$m to the continuum at $5.1-6.8\,\mu$m versus the ratio of
the strength of the $6.2\,\mu$m feature to the continuum at
$5.1-6.8\,\mu$m, \citet{Laurent00} attempted to quantify the
AGN/starburst contribution in the mid-IR. For source B1, their Fig.\,6
suggests that almost 40\,\% of the mid-IR emission could originate
from a very embedded AGN. The tell-tale sign would be the emission
observed via the ``hot continuum'' at $5\,\mu$m as illustrated in our
Fig.\,\ref{fig:bands}.  Even though the mid-IR evidence was not strong
enough to propose the presence of an AGN detection, their suggestion
was in fact proven true beyond any doubt with the recent Beppo-Sax
X-ray observations of Arp\,299 by \citet{Dellaceca02}.  These authors
explained the hard X-ray emission and spectral shape/luminosity of the
system with 
a buried Compton-thick AGN of intrinsic luminosity $L_{0.5-100\,{\rm
keV}} \simeq 1.9 \times 10^{43}\,{\rm erg\,s}^{-1}$, 
completely 
absorbed in the visible and in the IR ($A_{\rm V} > 1\,000\,$ 
magnitudes).
Their result, as well as
further work based on Chandra data which places the AGN in B1 \citep{Zezas03}, 
readily explains our inability to probe the
center of B1 in the optical and the UV.

An interesting point which remains to be explored in the future is
whether the AGN actually defines the location of a galactic nucleus in
NGC\,3690.  The reason for this question is that B2 is the most
extended source in the visible as seen in the HST WFPC2 ($0.59\,\mu$m)
images and the brightest one in the near-infrared, while B1 remains
unresolved or marginally resolved \citep{Alonso00}.  Moreover, unlike
B1, B2 does display a high CO index characteristic of a population of
late-type stars \citep{Satyapal99}.  Based on its near-IR colors and
its size at these wavelengths, B2 appears to be mainly composed of a
mix of old stars, gas and dust.  If this source is a relic of an older
single isolated starburst region outside the nucleus traced by B1, it
should have been spatially extended and of an extraordinary strength.

If, on the other hand, the nucleus of the galaxy is in B2, as
suggested by the near-IR colors, then this would imply that B1 harbors
not only massive star forming regions but also a ``runaway''
super-massive black hole (SMBH).  Did this SMBH originate from a
companion which was accreted in the system at an earlier stage and it
is now accreting gas becoming visible as an AGN? We do observe a high
concentration of gas outside -- but close to -- the B1 and B2 regions
of NGC\,3690 and their kinematics imply that they are tidally
connected with streamers originating from IC\,694 \citep{Casoli99}. 
\citet{Alonso00} though derived a dynamical mass for B1 ($M > 4 \times
10^{8}\,M_{\sun}$) similar to the one found in B2 ($M \sim 6 \times
10^{8}\,M_{\sun}$, from \citet{Shier96}) which would be hard to
explain if both of them had always been part of a single galaxy.  Our
mid-IR observations do not allow us to discriminate between the two
interpretations.  Only high spatial resolution information on the
kinematics of those components will help us shed some light to these
issues.

In any case, NGC\,3690 has been the host of extraordinary star-forming
events and exhibits a distinct mid-infrared spectral energy
distribution unlike any other in other extragalactic objects, 
even unambiguous 
AGNs such as 
NGC\,1068 \citep{Lefloch01} or Centaurus\,A \citep{Mirabel99}.

\section{Conclusions} 

Our mid-IR spectroscopy of Arp\,299 over the range 5 to $16\,\mu$m has
revealed new information on the properties of this intriguing
interacting system.  Based on our analysis, we conclude that:

\begin{enumerate}

\item Large quantities of dust are present in the system as indicated
by the hot continuum observed above $10\,\mu$m in all sources and by
the high extinction derived by our model.  We find that, assuming a
uniform foreground screen model, we can reproduce the mid-IR SED and
the $9.7\,\mu$m silicate absorption band with an $A_{\rm V}$ of $\sim
40$\,mag in the central region of IC\,694 (source A), $\sim 10$\,mag
for source C, and $\sim 50$\,mag for C$^\prime$, the dominant sources
in the galactic disks overlap region.  A best fit with an $A_{\rm
V}$ of $\sim 60$\,mag is derived in the central region of NGC\,3690
(source B1) using a homogeneous mixture of dust with radiating
sources.

\item The nuclear region of IC\,694 harbors enshrouded massive star
forming activity. No evidence for an active galactic nucleus is found
in our mid-IR spectra.

\item We confirm that C and C$^\prime$ are sites of recent starburst
events based on the elevated values of the [Ne{\sc iii}]15.6$\,\mu$m
flux and the high [Ne{\sc iii}]/[Ne{\sc ii}] ratio, which is similar
to those of classical young starbursts.

\item Source B is characterized by a unique mid-IR spectrum which is
dominated by a hot dust continuum extending shortward of $5\,\mu$m,
and which is the strongest of the Arp\,299 system.  Given the ample
presence of dust and the strength of the radiation field, this
continuum is consistent with the recent Chandra/Beppo-SAX results
which favor the presence of a Compton thick AGN at this location.

\end{enumerate}

\begin{acknowledgements}
We thank the HST team and PIs of the proposals who provided the WFPC2
NICMOS and FOC images made with the NASA/ESA Hubble Space Telescope
and obtained from data archives at the Space Telescope Science
Institute. STScI is operated by the Association of Universities for
Research in Astronomy, Inc. under the NASA contract NAS 5-26555.  VC
would like to thank F. Casoli (Obs. de Paris) for useful discussions
on the CO kinematics, A. Zezas (CfA/Harvard) for information on the
Chandra results prior to publication and acknowledges the
financial support of JPL contract 960803.
\end{acknowledgements} 
%

\end{document}